\documentclass[12pt]{article}
\usepackage{epsf}
\usepackage{epsfig}
\newcommand{\beq}{\begin{equation}}
\newcommand{\eeq}{\end{equation}}
\newcommand{\bea}{\begin{eqnarray}}
\newcommand{\eea}{\end{eqnarray}}

\newcommand{\gsim}{\lower.7ex\hbox{$
\;\stackrel{\textstyle>}{\sim}\;$}}
\newcommand{\lsim}{\lower.7ex\hbox{$
\;\stackrel{\textstyle<}{\sim}\;$}}

\newcommand{\eod}{\end{document}}

\newcommand{\gev}{\, {\rm GeV}}

\begin{document}
\thispagestyle{empty}
\vspace*{-22mm}

\begin{flushright}
UND-HEP-09-BIG\hspace*{.08em}05\\

\end{flushright}
\vspace*{1.3mm}

\begin{center}
{\Large {\bf
No Pain, No Gain -- On the Challenges and Promises \\
\vspace{2mm} of Charm Studies 
\footnote{Invited talk given at Charm09, Leimen (Germany), May 2009.} }}
\vspace*{10mm}

{\bf I.I.~Bigi} \\
\vspace{7mm}

 {\sl Department of Physics, University of Notre Dame du Lac}
\vspace*{-.8mm}\\
{\sl Notre Dame, IN 46556, USA}\\
{\sl email: ibigi@nd.edu} \\

\vspace*{10mm}

{\bf Abstract}\vspace*{-1.5mm}\\
\end{center}

\noindent
The observation of $D^0 - \bar D^0$ oscillations has left us in a tantalizing quandary concerning the theoretical 
interpretation: are they still compatible with the SM or might they require new dynamics (NP)? A comprehensive search for CP violation in $D$ decays should resolve the issue. Finding it 
should provide compelling evidence for the intervention of New Physics.  
While the absolute size of CP asymmetries will presumably be modest at best, the ratio of `signal' to `noise'  -- i.e. NP over SM contributions -- 
might well be larger for $D$ than $B$ transitions.  
A list of promising channels is provided, most of which should be observable in a hadronic environment. Yet to saturate the discovery potential for NP, we need a Superflavour Factory.   Valuable lessons can be obtained by analyzing three- and four-body final states.

\tableofcontents

\vspace{1.0cm}

\section{Prologue on Charm and its Uniqueness}

New Physics (NP) will in general induce flavour changing neutral currents (FCNC). The SM had to be crafted judiciously to have them greatly suppressed for strangeness; the weight of FCNC is then even more reduced for the up-type quarks $u$, $c$ and $t$. Yet NP scenarios could exhibit a very different pattern with FCNC being significantly more relevant for up-type quarks. Among those   
it is only the charm quark that allows the full range of probes for FCNC in general and for 
CP violation in particular. For top quarks do not hadronize \cite{RAPALLO} thus eliminating the occurrence of 
$T^0 - \bar T^0$ oscillations. Neutral pions etc. cannot oscillate, since they are their own antiparticles; furthermore CPT constraints are such that they rule out most CP asymmetries.  
Yet neutral charm mesons have been observed to oscillate: the world averages \cite{HFAG} based on data from BaBar \cite{BABAROSC}, Belle \cite{BELLEOSC} and CDF \cite{CDFOSC} read 
\bea
x_D = 0.0100^{+0.0024}_{-0.0026} \, &,& \;  y_D = 0.0076^{+0.0017}_{-0.0018}  \\
\frac{x_D^2 + y_D^2}{2} &\leq &(1.3 \pm 2.7)\cdot 10^{-4} 
\label{xyexp}
\eea
In the limit of (approximate) CP symmetry $x_D$, $y_D > 0$ implies the CP {\em even} state to be 
slightly heavier and shorter lived than the CP {\em odd} one (unlike for neutral kaons). 

While $D^0 - \bar D^0$ oscillations appear to have been established --- 
$(x_D,y_D) \neq (0,0)$ --- considerable uncertainty exists concerning both the absolute and 
relative sizes of $x_D$ and $y_D$; I will return to this point.  

I view the theoretical interpretation as unclear \cite{PETTALK}: while the SM can 
`naturally' generate $x_D \sim y_D \sim {\cal O}(10^{-3})$ \cite{URIDOSC,CICERONE}, one cannot rule out values as `large' 
as 0.01 \cite{FALK,CICERONE}. Despite the similar numerical estimates for $x_D$ and $y_D$ the underlying dynamics is of a very different nature: while $\Delta M_D$ is generated with {\em off}-shell intermediate states, $\Delta \Gamma_D$ is obtained from {\em on}-shell ones; 
$\Delta M_D$ can thus naturally be sensitive to  New Physics, which is unlikely for $\Delta \Gamma_D$. 

Finding $x_D \gg y_D \sim {\cal O}(10^{-3})$ would have represented strong prima facie evidence for 
the presence of NP. Such a scenario appears to have been ruled out. The present situation can instead be interpreted in two ways: (i) It is beyond our computational abilities to evaluate 
$\Delta M_D$ and $\Delta \Gamma_D$ accurately. (ii) It represents one example of nature being mischievous: $\Delta \Gamma_D$ is anomalously enhanced due to a violation of 
{\em local} quark-hadron duality caused by the proximity of hadronic thresholds 
\cite{CICERONE,DUA}; 
$\Delta M_D$ on the other hand is enhanced by NP over the value expected in the SM. 

My central point here is the following: while $x_D$, $y_D \sim 0.01$ might be produced by SM 
dynamics alone, {\em the observed size of $x_D$ might also contain a significant NP contribution}. 
I see no realistic way how this issue could be decided by theoretical means alone in the next few 
years. Yet there is a course of action to clarify and hopefully decide the issue, namely to conduct a comprehensive and dedicated study of CP invariance in charm decays. One 
can{\em not count} on NP creating large CP asymmetries in $D$ transitions, but its manifestations might be clearer here than in $B$ decays; for the SM creates much smaller 
"backgrounds"; i.e., it induces still much smaller effects: 
\beq 
\left[ \frac{\rm exp. \; NP\; signal}{\rm SM \; CP~"backgr."}\right]_{{\bf D}} > 
\left[ \frac{\rm exp. \; NP\; signal}{\rm SM \; CP~"backgr."}\right]_{{\bf B}}
\eeq
In summary: while the observed signal for $D^0 - \bar D^0$ oscillations represents a tactical draw in 
our struggle to reach beyond the SM, there is new promise for a strategic victory on the battleground 
of CP studies. It is the relative `dullness' of the SM vis-a-vis CP asymmetries in charm transitions that 
makes such searches promising. The observation of oscillations has widened the stage for NP to reveal itself.  

There is a {\em qualitative} analogy with the case of $B_s - \bar B_s$ oscillations: the observed rate as expressed through $\Delta M_{B_s}$ is fully consistent with SM predictions \cite{HFAG} -- within sizable theoretical 
uncertainties. The next challenge is to search for a time dependent CP asymmetry in 
$B_s \to \psi \phi$. For KM dynamics predicts \cite{BS80} a Cabibbo suppressed 
asymmetry $\sim 3$ \%. NP whose non-leading contribution to $\Delta  M_{B_s}$  
might be hiding behind the theoretical uncertainty in the SM prediction could enhance the CP asymmetry even by an order of magnitude thus becoming the 
{\em leading} effect {\em there}. 

\section{NP Scenarios and their Footprints}

\subsection{Fundamentals of CP Searches in Charm Decays}
SM predictions for $x_D$ and $y_D$ are unlikely to be refined significantly anytime soon. 
Even so determining $x_D$ and $y_D$ with good accuracy is motivated by pragmatic 
rather than quixotically noble reasons: a measurement of a presumably small time dependent CP asymmetry would be validated by reproducing values for $x_D$ and $y_D$ consistent with 
{\em independent} data. Those accurate values are also needed to identify the source(s) of an asymmetry, whether it is 
due to $\left| \frac{q}{p}\right| \neq 1$ or arg$\frac{q}{p}\bar \rho_f \neq 0$, as discussed later. 

Searching for manifestations of NP through CP studies is not a `wild goose' chase: Baryogenesis 
requires the intervention of NP {\em with} CP violation; it is not an unreasonable hope that such 
CP odd NP would leave its mark also somewhere else -- maybe even in charm transitions. One should note that a CP asymmetry is {\em linear} rather than quadratic in a NP amplitude, which enhances the 
sensitivity to small amplitudes. 

One drawback in searching for NP in charm decays is the fact that the latter occur already on the Cabibbo allowed level unlike for kaons and $B$ mesons, whose modes are Cabibbo and KM  suppressed, respectively. Since it is unlikely that NP in Cabibbo favoured channels would have escaped discovery, one has to search for it in significantly suppressed modes. Acquiring huge data sets is 
thus one essential requirement for our quest. 

Otherwise most 
{\em experimental}  features favour the observability of CP asymmetries in charm decays: 
(i) The effective branching ratios into pions, kaons and leptons for many relevant modes are relatively sizable. 
(ii) Final state interactions (FSI) needed to make direct CP violation observable in two-body final states are generally 
large \footnote{Our lack of theoretical control over final state interactions becomes a problem only when 
{\em interpreting} observations in terms of the {\em microscopic} parameters of the 
underlying dynamics.}. 
(iii) Flavour tagging can conveniently be done by the soft pions in $D^{*\pm} \to D/\bar D \pi^{\pm}$. 
(iv) Many nonleptonic  final states contain more than two pseudoscalar or 
one pseudoscalar and one vector meson. Final state {\em distributions} are then {\em non}trivial and can exhibit 
CP asymmetries, which can be studied through Dalitz plots and T odd moments. 
(v) The observation of 
$D^0 - \bar D^0$ oscillations greatly widens the stage for CP asymmetries, since it provides a second coherent, yet different amplitude, the weight of which changes with the time of decay. 

On the {\em phenomenological} side there are promising features as well. Since 
$D^0 \to \bar D^0$ transitions are so suppressed within the SM, they open a promising portal for 
NP to enter. $\Delta C=1$ nonleptonic modes occur on three Cabibbo levels -- allowed (CA), singly (SCS) and doubly Cabibbo suppressed (DCS) -- whose typical rates differ by tg$^2\theta _C$ and tg$^4\theta _C$, i.e. by one or two to three orders of magnitude, respectively. It is not unreasonable that NP could affect the decay amplitude for DCS and possibly even for SCS modes. Furthermore the SM provides not merely a classification -- it makes highly nontrivial, even if not  
precise predictions concerning CP asymmetries: no {\em direct} CP violation can occur in CA and DCS channels (except for final states containing $K_S$ [or $K_L$] mesons 
\cite{YAMA}); it can for SCS modes, but only on a tiny level, since the required weak phase is highly diluted to the tune of 
${\cal O}( {\rm tg}^4\theta _C) \leq 0.001$. In the SM model  
the oscillation amplitude is expected to carry a minute weak phase 
$\sim {\cal O}({\rm tg}^4\theta _C) \leq 0.001$ 
as a benchmark figure \cite{URIDOSC}. Even a NP contribution that is non-leading in $\Delta M_D$ can thus easily provide not only the leading source for CP violation, but even a sizable one.  

No CP asymmetry has been found in charm transitions. Let me specify that statement for indirect 
CP violation: 
\begin{itemize}
\item 
CP violation {\em in} oscillations is expressed by $|q/p| \neq 1$. The world average \cite{HFAG} reads: 
\beq 
\left| \frac{q_D}{p_D}\right| = 0.86 ^{+0.17}_{-0.15} \; . 
\label{qpexp}
\eeq
\item 
No CP asymmetry has been observed in $D^0 \to K^+K^-$, $\pi^+\pi^-$ \cite{PDG08}: 
\bea
\nonumber
A_{\rm CP}(K^+K^-) &=& (0.1 \pm 0.5) \% \\
A_{\rm CP}(\pi^+ \pi^-) &=& (0.0 \pm 0.5) \% 
\label{ACPKK}
\eea
\end{itemize}
As explained later I view these bounds as hardly telling: for the experimental sensitivity has only recently entered a domain, where one could `realistically hope' for an effect. 

\subsection{NP Models for Charm Dynamics}

The observation of $D^0 - \bar D^0$ oscillations has -- after some incubation period -- motivated theorists to analyze NP models  that could have an observable impact on  
$\Delta C=2$ transitions. Three complementary approaches have been 
tried: 

$\bullet$ 
One can review the `usual list of suspects' \cite{RENARD}; i.e., one analyzes different classes of NP models existing in the literature to analyze how large their impact can be \cite{PETTALK2}: it could be 
quite significant.  

$\bullet$  
One relies on an effective theory approach that reflects NP through an operator product expansion 
(OPE) containing higher dimensional operators construct from SM fields. One then infers a numerical value or bound for the coefficients of such operators from some observable. 
In principle this represents a model independent approach; in practice, however, one has to severely 
limit the number of operators included in the analysis: typically one considers just a single such operator with a specific Lorentz structure -- i.e., one does not allow for cancellations among different NP 
operators. Even so one can infer rough bounds for the scale characterizing NP; by invoking some 
global symmetry one can also obtain correlations among the impact of NP on different flavour sectors. A recent example is given in \cite{NIR1}, where connections between indirect CP violation in the 
strange and charm sectors are derived from assuming that the effective $\Delta S=2$ and 
$\Delta C=2$ four-quark operators induced by NP involve only quark doublets. As usual these connections are particularly significant for 1 TeV scale NP. SUSY can provide a straightforward 
dynamical implementation of such a scenario \cite{NIR1}.

$\bullet$ 
One can analyze models that are motivated by considerations quite {\em un}related to 
flavour dynamics. One such class of models is formed by little Higgs models with T parity. They are constructed to reconcile the 
{\em non}-observation of NP effects even in the quantum corrections to the electroweak parameters with the chance to find NP quanta at the LHC. This class of models is in general {\em not} of the Minimal Flavour Violation (MFV) variety.

\subsection{Littlest Higgs Model with T Parity}
\subsubsection{Basic Features}

Little Higgs models \cite{GENLH} 
are constructed to `delay the day of reckoning' for the gauge hierarchy problem. The Higgs boson appears as a Pseudo-Goldstone 
boson of a spontaneously broken global symmetry. This allows to keep the Higgs mass at most logarithmically divergent at the one-loop level, when its quadratic renormalization  
is arranged to vanish due to {\em bosonic} contributions from  a set of new heavy gauge bosons, a super-heavy cousin of the top quark etc. To accommodate the non-observation so far of NP 
contributions to the electroweak parameters even on the quantum level, a discrete symmetry called 
T parity is introduced \cite{LHTLIT,BLANKE}. To implement it, one has to introduce also mirror fermions --- one for each quark and lepton species -- that are odd under T parity. Flavour mixing in the mirror sector is described by two unitary matrices $V_{Hu}$ and $V_{Hd}$, parameterising the mirror quark
couplings to the SM up- and down-type quarks, respectively. Those matrices are related to each other 
by the CKM matrix: 
\beq\label{basic}
V_{Hu} = V_{Hd} V_{\rm CKM}^\dagger\,.
\label{CKMCON}
\eeq
With $V_{\rm CKM}^\dagger$ found to be close to the identity matrix, one expects several close 
connections between $K$ and $D$ decays, including with respect to CP violation. 
LHT thus provides a dynamical realization of the general ansatz made in \cite{NIR1}. 
One can  
express $V_{Hd}$ in terms of three mixing angles and three complex phases as suggested in 
\cite{Blanke:2006xr};  $V_{Hu}$, which shapes the LHT contributions to charm transitions, is then 
obtained from Eq.(\ref{basic}).

The charm analysis is undertaken assuming for the masses for the extra heavy gauge bosons 
\beq
M_{W_H, Z_H}= gf\sim 650\gev,\,  M_{A_H}=\frac{g'f}{\sqrt{5}}\sim 160\gev 
\eeq 
and the following range for the mirror fermions 
masses 
\beq
300\gev \le m_H^i \le 1000\gev\,.
\eeq
Such mass values are comfortably inside the reach for a direct detection at the LHC. 

\subsubsection{$D^0 - \bar D^0$ Oscillations}

Having found sets of LHT parameters consistent with the data {\em out}side charm dynamics (including 
those on $K$ and $B$ decays) we compute $M_{12}^D$ from them. As described in detail in 
\cite{BBBR} we found that the LHT contributions can yield a significant (or even leading) fraction of 
the observed value of $x_D = \Delta M_D/\Gamma_D$. More importantly for our discussion LHT can 
-- quite unlike the SM -- introduce sizable weak phases into $\Delta C = 2$ amplitudes. This feature 
can be expressed through the observables 
$\left| \frac{q}{p}\right|  \neq 1$ describing CP violation in $D^0 - \bar D^0$ oscillations 
and Im$\frac{q}{p}\bar \rho (f) \neq 0$ reflecting the interference between oscillations and 
(non-leptonic) decays \cite{CPBOOK2}. 
While $\left| \frac{q}{p}\right|  \neq 1$ can be probed most cleanly in neutral $D$ decays to `wrong-sign' 
leptons
\beq 
a_{\rm SL}(D^0) \equiv  \frac{\Gamma (D^0(t) \to \ell^-\bar\nu K^+) - 
\Gamma (\bar D^0 \to \ell^+\nu K^-)}
{\Gamma (D^0(t) \to \ell^-\bar\nu K^+) + \Gamma (\bar D^0 \to \ell^+\nu K^-)} 
= \frac{|q_D|^4 - |p_D|^4}{|q_D|^4 + |p_D|^4} \; , 
\eeq
it also affects non-leptonic modes in a prominent way together with 
Im$\frac{q}{p}\bar \rho (f) \neq 0$. The time dependent CP asymmetry generated by the latter can 
conveniently be expressed by 
\beq
\frac{\Gamma (D^0(t) \to f) - \Gamma (\bar D^0(t) \to f)}{\Gamma (D^0(t) \to f) + \Gamma (\bar D^0(t) \to f)}
\equiv S_f \frac{t}{2\bar \tau}
\label{SFDEF}
\eeq
for a CP eigenstate $f$. 
In the {\em absence of direct} CP violation these two quantities are necessarily related \cite{BBBR,NIR2}: 
\beq
S_f = - \eta_f \frac{x_D^2 + y_D^2}{y_D} a_{\rm SL}(D^0) 
\eeq
For the only source of CP violation is the relative phase between $\Gamma_{12}^D$ and $M_{12}^D$. 
LHT can certainly generate a much larger value for this weak phase than the SM; it can actually be 
quite sizable in particular when LHT contributes less than half the observed value for $\Delta M_D$. 

To be more specific: an extensive  sweep over the allowed LHT parameter space shows \cite{BBBR} 
\beq 
0.6 \lsim \left| \frac{q_D}{p_D}\right|^{\rm SM + LHT} \lsim 1.3
\label{QPLHT}
\eeq
to hold and thus 
\beq 
- 0.8 \lsim a_{SL}^{\rm SM+LHT}(D^0) \lsim + 0.3 \; . 
\label{ASLLHT}
\eeq
While the experimental information of Eq.(\ref{qpexp}) is fully consistent with CP invariance, it also allows for a large violation. Its two sigma band is similar to what LHT dynamics can typically produce. 
While we know already that the production of `wrong-sign' leptons is very low, see Eq.(\ref{xyexp}), 
their CP asymmetry could be very large -- unlike what we have in $K^0$, $B_d$ and $B_s$ decays. 

Furthermore we find that $S_f$ -- the time dependent CP asymmetry in nonleptonic $D^0$ decays 
-- can hardly exceed the about 1\% level, once we impose the experimental constraints on $x_D$, 
$y_D$ and $|q/p|$ \cite{BBBR}. This is the basis of my statement above that the present absence 
of an asymmetry -- Eq.(\ref{ACPKK}) -- is not very telling.  
The good news are that any improvement in experimental sensitivity can reveal a signal, and that 
CKM dynamics cannot produce more than ${\cal O}(10^{-5})$ effects. 

Intriguing connections emerge with other flavours \cite{BBBR}: 
\begin{itemize}
\item 
LHT can generate large 
deviations from CKM predictions  for $|q/p|_D$ and for the time-dependent CP asymmetry 
$S_{B_s \to \psi \phi}$ \cite{BS80} -- yet most likely {\em not for both}. 
\item
On the other hand large deviations can arise {\em simultaneously} for $|q/p|_D$ and for 
BR$(K_L\to \pi^0 \nu \bar \nu)$ -- even if $S_{B_s \to \psi \phi} < 0.05$ were found, i.e. 
a value practically indistinguishable from the CKM prediction. This is one example for the 
complementarity of the studies of $D$ and $K$ decays on one hand and of $B$ decays on the other 
in the search for NP.  

\end{itemize} 

LHT dynamics provides a new source for {\em direct} CP violation in Cabibbo suppressed $D$ channels through Penguin operators; their quantitative weight is being analyzed now.

\subsection{Cast of Candidate Channels}
\label{CAST}
Here I give a list of relevant channels, most of which -- but not all -- should be observable in hadronic collisions: 
\bea
\nonumber
D^0(t) &\to& K_SK^+K^-, \, K_S\pi^+\pi^- , \, K_S \eta ^{(\prime)} \\
\nonumber
D^0(t) &\to & l^- \bar \nu K^+ \\
\nonumber 
D^0(t) &\to& K^+K^-, \, \pi^+\pi^-, \, K^+\pi^- \\
\nonumber 
D^{\pm} &\to& K_S\pi^{\pm}, \; K_S K^{\pm} \\
\nonumber 
D &\to & 3\pi , \, K\bar K\pi , \, K^+\pi\pi \\ 
\nonumber 
D^0 &\to & K^+K^-\pi^+\pi^-, K^+K^-\mu^+\mu^-, K^+\pi^-\pi^+\pi^- \\
D^0 &\to& \mu^+\mu^- ,\, \gamma \gamma 
\eea
It should be noted that this is at best a representative list, not an exhaustive one. The dedicated 
reader is invited to come up with her/his personal favourite; in particular she/he can identify 
the corresponding $D_s$ modes. 

\subsection{List of Relevant Observables in Two-Body Modes}
\label{LIST} 

There are two a priori distinct portals for CP violation: it can enter via $\Delta C=1$ 
or $\Delta C=2$ dynamics referred to as direct and indirect CP violation, respectively. 
\begin{itemize}
\item 
{\em Direct} CP violation can reveal itself  
through 
a difference in the moduli of the $\Delta C=1$ decay amplitudes describing CP conjugate transitions: 
\beq 
|T(D \to f)| \neq |T(\bar D \to \bar f)| \; . 
\eeq
It requires the presence of two coherent amplitudes differing 
in both their weak as well as strong phases. 
\item 
The effects of $D^0 - \bar D^0$ oscillations on CP asymmetries can be expressed through 
\beq 
\frac{q_D}{p_D} = \sqrt{\frac{(M_{12}^D)^* - \frac{i}{2}(\Gamma_{12}^D)^*}
{M_{12}^D - \frac{i}{2}\Gamma_{12}^D}}
\eeq
Since $M_{12}^D$ and $\Gamma_{12}^D$ depend on the phase convention chosen for $\bar D^0$, neither of them nor $q_D/p_D$ can be observables by themselves. Yet $|q/p|_D$ is as already stated:  
$|q_D/p_D| \neq 1$ unequivocally describes {\em indirect}  CP violation in $D^0 - \bar D^0$ 
oscillations,  and it affects semileptonic and nonleptonic channels. 
\item 
{\em Tertium datur:} 
When a nonleptonic final state is common to $D^0$ and $\bar D^0$ decays -- in the simplest such 
cases it will be a CP eigenstate with parity $\eta_f$ --  it can exhibit a time-dependent asymmetry due to the 
interference between $D^0 - \bar D^0$ oscillations and $D$ decay. It can be 
expressed by $S_f$, see Eq.(\ref{SFDEF}); ignoring direct CP violation and with $x_D$, $y_D$ 
$\ll 1$ we can write down: 
\beq  
S_f= - \eta_f \left[ y_D \left(\left|\frac{q_D}{p_D}\right|- \left|\frac{p_D}{q_D}\right|\right)\cos2\varphi 
+ x_D \left(\left|\frac{q_D}{p_D}\right|+ \left|\frac{p_D}{q_D}\right|\right)\sin2\varphi \right]
\label{SMALLT}
\eeq
This expression shows why it is important to measure $x_D$ and $y_D$ as accurately as possible in an 
independent way: for one wants to determine whether an observed CP asymmetry is due to 
$|q| \neq |p|$ or $\varphi \neq 0$ or both. 

\item 
Oscillations can provide access to {\em direct} CP violation that might otherwise remain unobservable. 
Consider two different final states $f_1$ and $f_2$ that are CP eigenstates and thus 
common to $D^0$ and $\bar D^0$ decays; for example $f_1 = \phi K_S$ and $f_2=K^+K^-$. 
In the absence of direct CP violation one has 
\beq 
S_{f_1} \; = \; \eta_{f_1f_2} S_{f_2}
\eeq
with $\eta_{f_1f_2}$ denoting the relative CP parity of $f_1$ vs. $f_2$; it is -1 in the example given above. Any difference from this relation shows CP violation in at least one of the $\Delta C=1$ amplitudes. It should be noted that in the absence of oscillations this source of CP violation would be 
unobservable if the $D\to f_1$ and/or $D\to f_2$ amplitudes did not each contain two weak and 
two strong phases. 

\end{itemize}

I will list here channels that appear to be most promising for revealing such effects on various Cabibbo levels. Again no claim is made for this being a complete list -- the reader is invited to come up with her/his favourite modes.

\subsubsection{Two-Body Channels}
Searching for asymmetries in CA final states of neutral $D$ mesons represents a clean search 
for indirect CP violation, since NP affecting CA $\Delta C =1$ amplitudes should have been noticed by now. The theoretically simplest channels would be 
\beq 
D^0 \to K_S\pi^0, \, K_S\eta, \, K_S \eta^{\prime}
\eeq 
-- alas experimentally they 
are anything but simple. In a hadronic environment they seem to be close to impossible. The next best 
mode is
\beq 
D^0 \to K_S\phi \to K_S[K^+K^-]_{\phi} \; , 
\eeq
which is given by a single 
isospin amplitude. The {\em strong} phase thus drops out from the ratio 
$\bar \rho_{K_S\phi} \equiv \frac{T(\bar D^0 \to K_S\phi)}{T(D^0 \to K_S\phi)}$, while their 
tiny SM weak 
phase can be ignored for the time being. We have here a {\em qualitative}  analogy to $B_d \to \psi K_S$. The effect will be much smaller of course with 
much slower oscillations, and a priori one cannot ignore the impact of $y_D \neq 0$ and 
$\left| \frac{q_D}{p_D} \right| \neq 1$.

Extracting $D^0 \to K_S \phi$ from $D^0 \to K_SK^+K^-$ is not trivial. Among other challenges one has to distinguish it 
from $D^0 \to K_Sf^0$. For the CP parities of 
$K_Sf^0$ and $K_S\phi$ are opposite. Therefore these final states would have to exhibit CP asymmetries of equal size, yet opposite sign. Ultimately one will perform a 
CP analysis of the full Dalitz plot for $K_SK^+K^-$ -- a topic I will address below.  

The mode 
\beq
D^{\pm} \to K_S \pi^{\pm} 
\eeq
at first sight appears to be a CA mode. However that final state can be reached also through a DCS amplitude -- 
$D^{\pm} \to \overline K^0 \pi^{\pm} / K^0 \pi^{\pm} \to K_S \pi^{\pm}$ -- and its interference with the 
CA amplitude provides a non-negligible contribution. With{\em out} NP there are already two sources for a direct CP asymmetry \cite{YAMA}: (i) The interference between 
$D^{\pm} \to \overline K^0\pi^{\pm} \to K_S\pi^{\pm}$ and 
$D^{\pm} \to K^0\pi^{\pm} \to K_S\pi^{\pm}$ generates an asymmetry 
$\sim {\cal O}({\rm tg}^6\theta_C ) \sim 10^{-4}$. (ii) The CP impurity in the $K_S$ wave function 
induces a larger asymmetry: 
\beq 
\frac{\Gamma (D^+\to K_S\pi^+) - \Gamma (D^-\to K_S\pi^-)}
{\Gamma (D^+\to K_S\pi^+) + \Gamma (D^-\to K_S\pi^-)} 
\simeq 
\frac{|q_K|^2 - |p_K|^2}{|q_K|^2 + |p_K|^2} 
\simeq - (3.32 \pm 0.06) \cdot 10^{-3}
\eeq
Any deviation from this accurate prediction would be due to an intervention by NP, presumably 
entering through the DCS amplitude. 

The situation becomes more complex for Cabibbo suppressed channels. For CKM dynamics can 
already induce 
CP asymmetries -- albeit highly diluted ones to the tune of $10^{-3}$ or less -- since two different 
isospin amplitudes contribute. That opens the door wider for NP to enhance an asymmetry over its tiny 
SM expectation, even when it yields no more than a non-leading contribution to the rate. 

Prime examples for promising channels are 
\beq 
D^0 \to K^+K^-, \; \pi^+\pi^-
\eeq
where now direct as well as indirect CP violation can arise. In the absence of the former the time dependent asymmetry has to be the same for both channels, since it is driven by the oscillations 
common to both modes. While no asymmetry has been observed there yet -- see Eq.(\ref{ACPKK}) -- 
a signal could hardly have emerged on that level, but any improvement in the experimental sensitivity for  
$D^0(t) \to K^+K^-, \, \pi^+\pi^-$ constrains NP scenarios -- or could reveal them 
\cite{GKN}.

Since $D^{\pm} \to \pi^{\pm}\pi^0$ leads to a pure isospin-two final state, CPT and isospin symmetries 
combine to severely limit its   
CP asymmetry. The situation is more promising for 
\beq 
D^{\pm} \to K^{\pm}K_S 
\eeq
with $D^{\pm} \to \pi^{\pm}\eta^{(\prime)}$ providing the compensating asymmetry to satisfy CPT constraints.

Some DCS contributions have already been considered, when they enter through the `backdoor' 
of final states containing a $K_S$ (or $K_L$). A promising pure DCS channel 
is \cite{BIBERK,NIRETAL}: 
\beq 
D^0 \to K^+\pi^-
\eeq
\begin{itemize}
\item 
For it allows to track both direct and indirect CP violation and separate it through analyzing how the asymmetries evolve with the time of decay. 
\item 
The SM amplitude being DCS is significantly reduced by tg$^2\theta_C \sim 1/20$. 

\end{itemize}
It should be noted that sources of indirect and direct CP violation could be quite unrelated to each 
other \cite{GAO}.

\section{Final State Distributions}
\subsection{Dalitz Plot Studies}

Final states with three pseudoscalar mesons can be treated in a `Catholic' style: the Dalitz 
plot provides a single path to `heaven'. The challenge we face here can be summarized as follows: we look for probably smallish asymmetries in subdomains of the Dalitz plot, which is shaped by 
nonperturbative dynamics. While a proper Dalitz analysis requires a considerable `overhead' in setting it up, it offers -- like 
T {\em odd} correlations addressed next -- valuable `pay-offs': 
\begin{itemize}
\item 
Local asymmetries are bound to be larger than integrated ones. 
\item 
There are correlations in a Dalitz plot that a proper analysis has to exhibit. Such correlations 
provide us with powerful validation tools in particular for smallish effects. 
\item 
CP asymmetries in a Dalitz plot can provide us with information about the 
underlying operators -- whether the Lagrangian is built from the products of spin-zero or spin-one operators, whether it contains 
mixed-chirality quark fields -- that are not revealed in two-body modes. 

\end{itemize}
Constructing and tuning a model for a full Dalitz plot description requires very large statistics -- and even then one can{\em not} count on a {\em unique} model. Let me list just two reasons for ambiguities: (i) The 
non-resonant contributions are usually assumed to be flat across the Dalitz plot; yet it is merely 
mathematical simplicity rather than a dynamical insight that suggests such an ansatz. Assuming a 
non-uniform distribution -- reflecting, say, a threshold enhancement -- can have a considerable impact on what a fit yields for the resonant contributions. 
(ii)  While pseudoscalar and vector resonances can adequately be described by Breit-Wigner 
excitation curves, chiral dynamics tell us this is not the case for scalar resonances like the sigma or 
$\kappa$. It is a rather subtle dynamical question how those are to be described, and it will vary from 
channel to channel. 

Thus I see a two-fold challenge in front of us: 
\begin{itemize}
\item 
Some dedicated theoretical effort has to be made to refine our tools for Dalitz plot descriptions 
\cite{BHM}. 
\item 
Even when maintaining that a description in terms of 
specific resonant and nonresonant amplitudes is the ultimate method for extracting all dynamical 
information from the data, it makes sense to develop alternative methods of analysis that are more 
robust and less model dependent, even if they cannot provide us with the full dynamical information. 
Such methods might enable us to draw firm, although less complete conclusions from more  limited 
data sets. They could reveal more quickly the existence of a CP asymmetry, which in the case of charm 
decays might be tantamount to establish the intervention of NP, and maybe even localize the 
sub-domain of the Dalitz plot, where the main source of the asymmetry resides. At the very least 
it would provide us with diagnostics concerning the critical domains for the Dalitz plot models. 

\end{itemize}
Such alternative methods likely revolve issues of pattern recognition. There we can learn a lot from astronomers. They regularly face the problem of searching for something they do not quite know what it is at a priori unknown locations and having to deal with background sources that are all too often not really understood. While this sounds like a hopeless proposition, astronomers have actually been quite successful in overcoming these odds. Inspired by astronomers an intriguing suggestion has been made \cite{MIRANDA}: rather than search for the customary asymmetry 
$(N- \bar N)/(N + \bar N)$ in particle vs. anti-particle populations $N$ and $\bar N$, respectively, analyze 
\beq 
\Sigma \equiv \frac{N- \bar N}{\sqrt{N + \bar N}} 
\label{SIGVAR}
\eeq
It corresponds to standard procedure in astronomy -- suggested in 1983 for gamma ray 
astronomy \cite{ASTRO} and now adopted also by the Auger collaboration --  when comparing on- vs. off-source intensity. For a Poissonian distribution the standard deviation can be written as 
$\sigma = \frac{N_{\rm on}- \alpha N_{\rm off}}{\sqrt{N_{\rm on} + \alpha N_{\rm off}}}$. In the pilot 
study of \cite{MIRANDA} we have analyzed a few scenarios for Monte Carlo generated 
$B^{\pm}\to K^{\pm}\pi^+\pi^-$ 
and $D^{\pm} \to \pi^{\pm}\pi^+\pi^-$ decays, where we had seeded just a single source for 
CP violation. Using the variable $\Sigma$ of  Eq.(\ref{SIGVAR}) we could extract robust signals for the 
existence of a CP asymmetry and identify correctly the approximate location of the seeded 
asymmetry. More case studies are under consideration, including those involving time 
dependent Dalitz plots due to oscillations. It would be most desirable to test this method with real data 
to get a fuller evaluation of its potential.

This is just one possible example for how we can learn from our astronomer colleagues -- I am sure there will be more under the motto: "Copying is the highest form of flattery".

\subsection{T {\em Odd} Correlations}

Going beyond three-body final states one has to deal with a `Calvinist' situation. {\em A priori} there are several paths to heaven, and heaven's blessing is revealed {\em a posteriori} by the success of one's 
efforts; i.e. which distribution will provide the clearest CP asymmetry depends on the specifics of the underlying dynamics. This is good and bad in a complementary way: `bad' in the sense that in a general 
search for NP one has to analyze several distributions; `good', because once one has found a distribution -- or a moment of such -- that reveals an asymmetry, then its form can tell us something 
important about the underlying NP; or one can design the optimal observable if one searches for a very specific form of NP. 

Since under T both momenta $\vec p$ and spin vectors $\vec s$ change sign, the most elementary T odd 
moments are given by expectation values of triple correlations like 
\beq 
\langle \vec p_1\cdot (\vec p_2 \times \vec p_3 \rangle \; \; \; {\rm and/or} \; \; \; 
\langle \vec s\cdot (\vec p_1 \times \vec p_2 \rangle
\eeq
Unless one has access to spin vectors one obviously needs at least a four-body final state for a 
T odd moment. 

There is a subtle, yet important distinction between T odd and, say, P odd moments. Observing a 
P odd moment unequivocally establishes P violation unlike for the case of a T odd moment. 
For the latter can be generated also with T invariant dynamics if one goes beyond lowest order; 
i.e. FSI can {\em fake} a T violation \cite{CPBOOK2}. This complication is due to time reversal being described by 
{\em anti}linear transformations. While FSI are a necessary evil for CP asymmetries to emerge in partial rates, they can be a nuisance for T odd effects: while they are not needed, they can fake an effect.  

There are two ways to deal with this interpretative challenge: (i) One can attempt to estimate the order of magnitude of such FSI effects. (ii) One can compare T odd moments in  CP conjugate  
decays of particles and antiparticles: If they are not equal in magnitude, yet opposite in sign, 
T invariance is broken, since CP transformations are linear. 

The simplest cases are provided by \cite{KAON,CICERONE} 
\bea
D^0 \to K^+K^- \pi^+\pi^-      \;  & vs. &\; \bar D^0 \to K^+K^- \pi^+\pi^-   \\
D^0 \to K^+K^- \mu^+\mu^-      \; & vs. &\;  \bar D^0 \to K^+K^- \mu^+\mu^-   \\
D^+ \to K^+K_S \pi^+\pi^-      \;  & vs. &  \;  D^- \to K^-K_S \pi^+\pi^-   
\eea
since all particles in the final state are distinct. It should be emphasized again that these are merely 
the most straightforward channels. A pioneering analysis of such correlations has been undertaken 
by the FOCUS Collab. \cite{FOCUS} and is now under study by BaBar \cite{ANTIMO}. 

One can also analyze, say, $D^0 \to \pi^+\pi^- \pi^+\pi^-$ 
and use selection criteria like the energies of the pions.  It makes sense to analyze T odd 
moments for neutral $D$ decays as a function of the (proper) time of decay, since oscillations 
affect the relative weight of different contributions. 

Likewise one can use different kinematic variables to form T odd moments. One can measure the 
azimuthal angle between the $K \bar K$ and the $\pi^+\pi^-$ or $\mu^+\mu^-$ planes and search for a 
forward-backward asymmetry in it -- in analogy to what has been done for 
$K_L \to \pi^+\pi^-e^+e^-$ \cite{SEHGAL}. Without a specific model for the underlying CP odd dynamics one cannot decide a priori which correlations are most sensitive to NP dynamics. 

\section{Benchmark Goals}

Viable NP scenarios could produce CP asymmetries close to the present experimental bounds, but 
not much higher. To have a `fighting' chance to find an effect, one should strive to reach 

$\bullet$ 
the ${\cal O}(10^{-4})$ [${\cal O}(10^{-3})$] level for time-dependent CP rate asymmetries in 
$D^0 \to K^+K^-$, $\pi^+\pi^-$, $K_S\rho^0$, $K_S\phi$ [$D^0 \to K^+\pi^-$]; 

$\bullet$ 
direct CP asymmetries in partial widths down to ${\cal O}(10^{-3})$ in $D\to K_S\pi$ and  in 
singly Cabibbo suppressed modes and down to ${\cal O}(10^{-2})$ in doubly Cabibbo suppressed 
modes; 

$\bullet$
the ${\cal O}(10^{-3})$ level in Dalitz asymmetries and T odd moments.

\section{On Rare Charm Decays}

There seems to be general agreement that studying $D \to \gamma X$ etc. is very unlikely to allow establishing the presence of NP because of uncertainties due to long distance dynamics 
\cite{BURD}. I am concerned that the same strong caveat applies also to $D \to l^+l^- X$. 

The story is more promising for $D^0 \to \mu^+\mu^-$. While its rate suffers greatly from helicity suppression and the 
need for weak annihilation -- even the first factor is almost model independent -- it is easier to 
interpret. In the SM the rate is estimated to be greatly dominated by long-distance dynamics -- yet on a very tiny level \cite{BURD}: 
\beq
{\rm BR}(D^0 \to \mu^+\mu^-)_{\rm SM} \simeq {\rm BR}(D^0 \to \mu^+\mu^-)_{\rm LD}  
\simeq  
3\cdot 10^{-5} \times {\rm BR}(D^0 \to \gamma\gamma )_{\rm SM} 
\eeq
With the SM contribution to $D^0 \to \gamma \gamma$ again being dominated by long-distance 
forces \cite{BURD} 
\beq
{\rm BR}(D^0 \to \gamma\gamma )_{\rm SM} \simeq {\rm BR}(D^0 \to \gamma\gamma )_{\rm LD} 
\sim (1 \pm 0.5)\cdot 10^{-8} \; , 
\eeq
one infers 
\beq 
{\rm BR}(D^0 \to \mu^+\mu^-)_{\rm SM} \sim 3 \cdot 10^{-13} 
\eeq
to be compared with the present bounds 
\bea 
\label{MUMU}
{\rm BR}(D^0 \to \mu^+\mu^-) &\leq& 5.3 \cdot 10^{-7}  \\ 
\label{2GAMMA}
{\rm BR}(D^0 \to \gamma \gamma) &\leq& 2.7 \cdot 10^{-5}  .  
\eea
The bound of Eq.(\ref{2GAMMA}) implies a bound of $10^{-9}$ in Eq.(\ref{MUMU}) -- i.e., a 
much tighter one. 
In either case there is a rather wide window of opportunity for discovering 
NP in $D^0 \to \mu^+\mu^-$. As pointed out in \cite{PETROV2} in several NP models there is actually a relatively tight 
connection between the NP contributions to BR$(D^0 \to \mu^+\mu^-)$ and 
$\Delta M_D/\Gamma_D$. 

Specifically LHT makes short-distance contributions to $D^0 \to \mu^+\mu^-$ and 
$D^0 \to \gamma \gamma$ that can be calculated in a straightforward way as 
a function of viable LHT parameters. Their size is under active study now. No matter what 
drives $D^0 \to \gamma \gamma$ - whether it is from short or long distance dynamics -- it 
provides a long distance contribution to $D^0 \to \mu^+\mu^-$. For a proper interpretation of these 
rare $D$ decays it is thus important to search for $D^0 \to \gamma \gamma$ with as high a 
sensitivity as possible.

\section{On Theoretical Guidance}
\label{GUID} 

To say that theoretical predictions have not always been on the mark, in particular when nonperturbative forces are involved, is putting it delicately. Yet this does not justify immediate rejection of theoretical advice -- it merely points to the need for some healthy skepticism.  

It is my considered judgment that the `hadronic community' has acquired a great deal of expertise and 
accumulated a wealth of information on low energy hadronic interactions. Yet this knowledge 
has hardly migrated to the heavy flavour community; it could be put to excellent and urgently needed use in studies of charm and beauty decays with unprecedented statistics. It is high time that 
a broader bridge is built between the two communities -- and we intend to do just that \cite{BHM}.

\subsection{`Theoretical Engineering'}
\label{THEORENG}

When describing (quasi) two-body channels of the $D \to PP$, $PV$ type we can make an intelligent use of measured partial rates to get to a reasonably reliable theoretical description of CP asymmetries there. 
On each Cabibbo level one expresses the total transition operator in terms of 
`elementary' $\Delta C = 1$ operators whose coefficients are computed from the known CKM factors and QCD radiative corrections. One makes a judicious choice of which such $\Delta C = 1$ operators to include -- corresponding to internal and external $W$ emission with or without interference, weak annihilation and Penguin contributions if possible. When evaluating the corresponding amplitudes one leaves the magnitude of the appropriate strong matrix elements and the values of their FSI phases open. From 
fitting such expressions to a comprehensive set of high statistics data one infers values for these a priori unknowns. The reliability of such extractions rests on the degree of over-constraints one has achieved including cross referencing those numbers against each other using $SU(3)_{fl}$ relations etc. The ability to include also channels with (multi-)neutrals is obviously of essential value here; 
such measurements belong to the domain of $e^+e^-$ $\tau$-charm factories like BESIII 
\cite{BESBOOK}.  

\subsection{On CPT Constraints}
\label{CPT}

CPT symmetry provides more constraints than just equality of masses and lifetimes of particles and antiparticles \cite{CPBOOK2}. For it tells us that the widths for {\em sub}classes of transitions have to be the same. For 
simplicity consider a toy model where the $D$ meson can decay only into two classes of final states 
$E=\{ e_i, i=1,...,n \}$ and $F=\{f_j, j=1,...,m \}$ with the strong interactions allowing members of the 
class $E$ to rescatter into each other and likewise for class $F$, but {\em no} rescattering possible 
{\em between} classes $E$ and $F$. Then CPT symmetry tells us partial width asymmetries 
{\em summed} over class $E$ already have to vanish and likewise for class $F$. This CPT `filter' can hardly be of any 
practical use for $B$ decays with their multitude of channels on vastly different CKM levels. 
Yet it might provide nontrivial validation checks for $D$ decays with their considerably fewer channels, where quasi-{\em elastic} unitarity could conceivably hold in a semiquantitative way. 

Penguins despite their poor reputation in flavour dynamics -- as expressed through the 
all too often heard "penguin pollution" -- are rather smart beings. When Penguin diagrams are invoked 
to generate the FSI required for a direct CP asymmetry, examining the light quarks in their loops will tell you in which class of channels the compensating asymmetries have to arise. To cite a simple example: such considerations suggest that a 
direct CP asymmetry in $D^0 \to K^+ K^-$ is compensated mainly by an asymmetry in 
$D^0 \to \pi^+ \pi^-$.  Finding these balancing effects would validate the observation of a presumably small asymmetry.

\subsection{On Relating Direct and Indirect Searches for New Physics}
\label{NPSCEN}

Looking for NP inducing CP violation in charm decays represents `hypothesis-generating' research -- 
similar to the {\em present} situation in $B$ physics. Once, say, SUSY is found in high 
$p_{\perp}$  collisions at the LHC, future studies of $B$ decays would be of the 
`hypothesis-probing' variety; this would be analogous to the situation about fifteen years ago, when the $e^+e^-$ $B$ factories were approved. Finding direct evidence for LHT models might turn the same trick for the detailed study of 
$D$ decays.

\section{Summary and Outlook}
\label{OUT}

There is general conviction that $D^0 - \bar D^0$ oscillations have been observed:  
$(x_D,y_D) \neq (0,0)$. However their theoretical interpretation is rather ambiguous: 
SM dynamics might generate the whole effect or the major part of it or only a minor part. 
Deciding this issue on theoretical grounds would require a breakthrough in our computational 
abilities. A comprehensive and detailed program of CP studies in charm transitions can presumably 
decide the issue: It might establish the intervention of NP; for even if it provided 
only a non-leading contribution to $\Delta M_D$, it would quite possibly represent the leading source of 
CP asymmetries due to the `dullness' of SM CP phenomenology. The present absence of a signal 
CP violation is 
not very telling. For future studies we need to know the relative size of 
$x_D$ and $y_D$ as best as possible. 

`Realistically' one cannot hope for much more than ${\cal O}(10^{-3})$ effects. Thus we have to learn to 
exploit the statistical `muscle' of LHCb and control systematics. Asymmetries in final state distributions as analyzed through Dalitz studies and T odd correlations offer several advantages: differential asymmetries could be considerably larger than integrated ones; internal cross checks provide powerful tools to deal with systematics; they can provide us with novel clues about the nature of the intervening New Physics. 
On the 
theory side we can expect a positive learning curve for theorists, yet should not expect  
miracles. What we can count on is that precise data will prompt some theorists to take on the 
challenge of developing an adequate description. 

My plea for more dedicated charm studies is not merely a repetition of a "Ceterum censeo fascinum 
esse studiandum" ("Moreover I advise that charm has to be studied"). The field of charm studies has achieved a qualitatively new level of maturity and promise through the observation of 
$D^0 - \bar D^0$ oscillations and the `awakening' this discovery has prompted in the theory community 
concerning the possibility of NP scenarios leaving their clearest footprint in charm decays.  

\begin{figure}[ht]
\begin{center}
\epsfig{bbllx=-4cm,bblly= 1.5cm,bburx=22cm,bbury=28cm,
height=11truecm, width=12truecm,
        figure=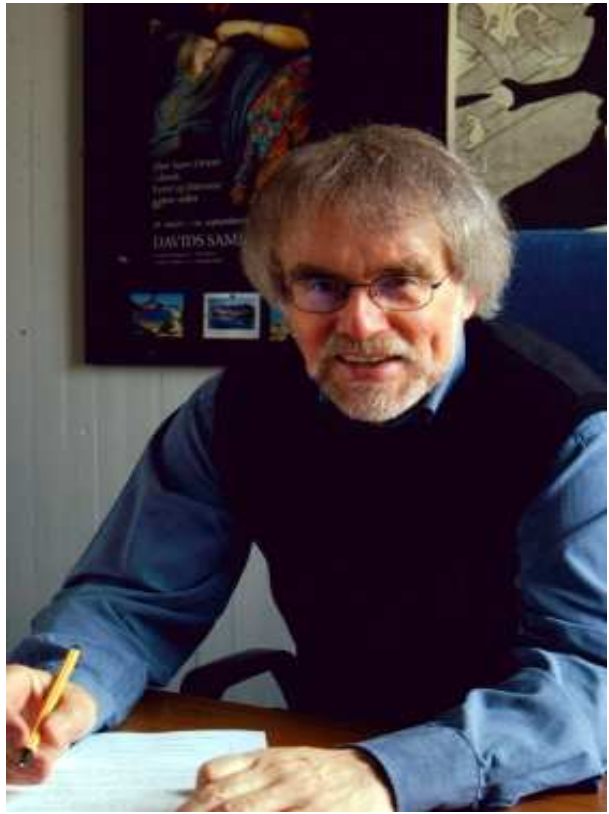}
\caption{A Polish Gentleman boasting now (and forever ?) a superbly-groomed beard.}
\label{BB}  
\end{center}
\end{figure}

Andrzej Buras has authorized me to make the following statement:
He is willing to bet his beard that LHT models would lead to
observable CP violation in D decays! Your studies of charm decays will thus have significant impact 
irrespective of their outcome: {\em If} you find CP violation, you have most likely discovered the 
intervention of NP. {\em If not}, you will create an even more visible impact as you can imagine 
from Fig. \ref{BB}!  

\section{Epilogue: Bismarck's Dictum}

Bismarck -- a statesman with firm goals, yet not given to moral qualms about how to achieve those -- 
once declared: "... the role of the statesman is to grab the mantle of history when he feels it passing by ...". Likewise it is the task of the physicist to make the greatest use of a special gift from Nature. 
$D^0 - \bar D^0$ oscillations are such a gift. Therefore it is your duty to make the best use of it -- and there is fame within your grasp!

\vspace{0.5cm}

{\bf Acknowledgments:} I am grateful to Klaus Peters and his team for organizing this fine conference in 
such a nice `Tuscan' environment. I have benefitted from numerous discussions with my colleagues, 
in particular J. Appel, I. Bediaga, Ch. Hanhart, Th. Mannel, J. Miranda, A. Reis and G. Wilkinson. This work was supported by the NSF under the grant number PHY-0807959.

\vspace{4mm}



\begin{thebibliography}{99}
\bibitem{BHM}
Ch. Hanhart, Ulf Mei\ss ner and I are founding a working group to deal with this challenge 
in a systematic way.
\end{thebibliography}
\end{document}